\documentclass{article}
\usepackage{spconf,amsmath,graphicx,url,amssymb}
\usepackage{subfigure}
\usepackage{graphicx}
\usepackage{color}
\usepackage{enumitem}

\title{Supervised Chorus Detection for Popular Music Using Convolutional Neural Network and Multi-task Learning}
\name{Ju-Chiang Wang, Jordan B.L. Smith, Jitong Chen,  Xuchen Song, and Yuxuan Wang}
\address{ByteDance\\ {\small\tt \{ju-chiang.wang, jordan.smith, 
chenjitong.1, xuchen.song, wangyuxuan.11\}@bytedance.com}}
\begin{document}
%
\maketitle
\begin{abstract}

This paper presents a novel supervised approach to detecting the chorus segments in popular music. Traditional approaches to this task are mostly unsupervised, with pipelines designed to target some quality that is assumed to define ``chorusness,'' which usually means seeking the loudest or most frequently repeated sections. We propose to use a convolutional neural network with a multi-task learning objective, which simultaneously fits two temporal activation curves: one indicating ``chorusness'' as a function of time, and the other the location of the boundaries. We also propose a post-processing method that jointly takes into account the chorus and boundary predictions to produce binary output. In experiments using three datasets, we compare our system to a set of public implementations of other segmentation and chorus-detection algorithms, and find our approach performs significantly better.

\end{abstract}


\section{Introduction}\label{sec:introduction}

Verse-chorus song form is a very common structure for popular music.
In it, verses alternate with choruses, with the lyrics of the verses varying and the choruses repeating more strictly and more frequently.
The authors of~\cite{vanbalen2013analysis}
cite other generalizations used to define choruses, including that they are the `most prominent' and  `most catchy' sections of a piece.
These traits make it desirable to detect choruses automatically,
whether for generating ``thumbnails''~\cite{bartsch2001catch, chai2003thumbnailing, muller2012robust},
for finding the emotional ``highlights'' of a piece~\cite{huang2018pop},
or for enabling convenient navigation based on the song structure~\cite{goto2003smartmusickiosk}.

However, most previous approaches to chorus detection and thumbnailing~\cite{goto2006chorus,eronen2007chorus,bartsch2001catch,chai2003thumbnailing} are unsupervised.
They begin with an observation about what typefies chorus sections, and search for them on this basis: e.g., finding the loudest, most frequently repeated, and/or the most homogenous section.
Since the definition of `chorus' is a generalization that does not apply in all cases, even a perfectly-designed system of this type will fail to detect the chorus in many songs.
A better approach may be to let a model \emph{learn} what defines `chorusness' from labeled examples;
this would allow a system to leverage the timbral and spectral features identified by~\cite{vanbalen2013analysis} in a study of what acoustic features differentiate choruses.

This approach, when applied to the related task of music boundary detection by~\cite{ullrich2014_ismir}, led to a huge leap in the state of the art. Prior segmentation algorithms would generally focus on a definable proxy task (e.g., detecting points of change or onsets of repetitions),
assisted by sensible heuristics (e.g., rounding boundary estimates to the nearest downbeat). A convolutional neural network (CNN) is trained to detect whether the center of a 16-second input is a boundary. When post-processed with an appropriate threshold, \cite{ullrich2014_ismir} demonstrated a 10\% improvement in f-measure over the state of the art.

We propose a similar approach: train a neural network to predict the ``chorusness'' of an excerpt directly from the audio, and without the context of the rest of the song.
We train a binary classifier to predict the ``chorusness'' of each point in a window, and slide this window throughout the song to obtain a chorus probability curve.
However, this leaves the problem of finding an appropriate threshold for post-processing.
To ease this, we propose to jointly model the chorus activation and boundary activation curves, so that the loss on the signals around the boundaries is naturally emphasized. At the inference phase, it also eases the process of converting the raw probability curve to a binary output for a song.


Chorus detection is clearly related to two tasks with a long tradition of MIR research: thumbnailing and music structure analysis (MSA)~\cite{muller2015a}.
The objective of thumbnailing is to find a short excerpt of a song that would be an effective preview.
However, there is no definition of what makes a good preview; \cite{chai2003thumbnailing} cited several. In practice, thumbnailing systems are evaluated by testing how often they select all or part of a chorus~\cite{bartsch2001catch}, or whichever segment is repeated most often~\cite{muller2012robust}.
Recently, \cite{huang2018pop} proposed a novel, related objective---to find the emotional highlights of pop songs---and evaluated their system based on whether it captured the choruses, which were assumed to correspond to the highlights, but their system used a neural network trained to detect emotion, not choruses.

In music structure analysis, it is assumed that one family of segments corresponds to the chorus, but predicting which one is only rarely attempted. We are aware of three prior systems:
\cite{maddage2004content}, who assumed a highly restricted template for song structures and used heuristics to predict labels; \cite{paulus2010improving}, who paired a standard structure analysis system with an HMM trained to label the sections; and \cite{shibata2020music}, published very recently, who proposed a hierarchical generative model (with section parts generating chord progressions, and these in turn generating observed feature sequences). This last model benefits from supervision, but still relies on a hand-set strategy of detecting homogeneity and repetitions, based on handcrafted features (chroma and MFCCs).

The lack of attention paid to chorus detection may be due to the difficulty of obtaining sufficient training data.
SALAMI~\cite{smith2011design} contains 1446 songs, but these come from diverse genres, so it may be difficult to learn a coherent notion of ``chorusness'' from it.
Introduced in 2019, the Harmonix Set~\cite{NietoISMIR2019} contains 912 songs, 888 with ``chorus'' sections; it is the most frequent label, with over 3100 choruses altogether, which is 41\% more than the ``verse'' instances.
We also have the annotated chorus locations for an internal dataset (denoted as \emph{In-House}) of 2480 Asian pop songs.
We use these three sources to train or evaluate our system.
Since the data sources all have different properties, we investigate the cross-dataset performance of our system.


In addition to the usefulness of detecting choruses for other applications, the annotations of choruses (that we depend on) seem more reliable than for other sections.
In SALAMI, we observed that if one annotator perceives a segment starting at time $t$, there is a 66\% chance that the other annotator placed a boundary at the same time (within 0.5 seconds)---but this probability rises to 78\% if the boundary marks the start of a `chorus'. This greater agreement could be the result of choruses having more salient beginnings than other section types~\cite{bruderer2009perception}. 
Therefore, the reliability of the annotations makes a supervised system more feasible.

\section{Proposed Approach}\label{sec:method}

\begin{figure}
\centering
\includegraphics[width=\columnwidth]{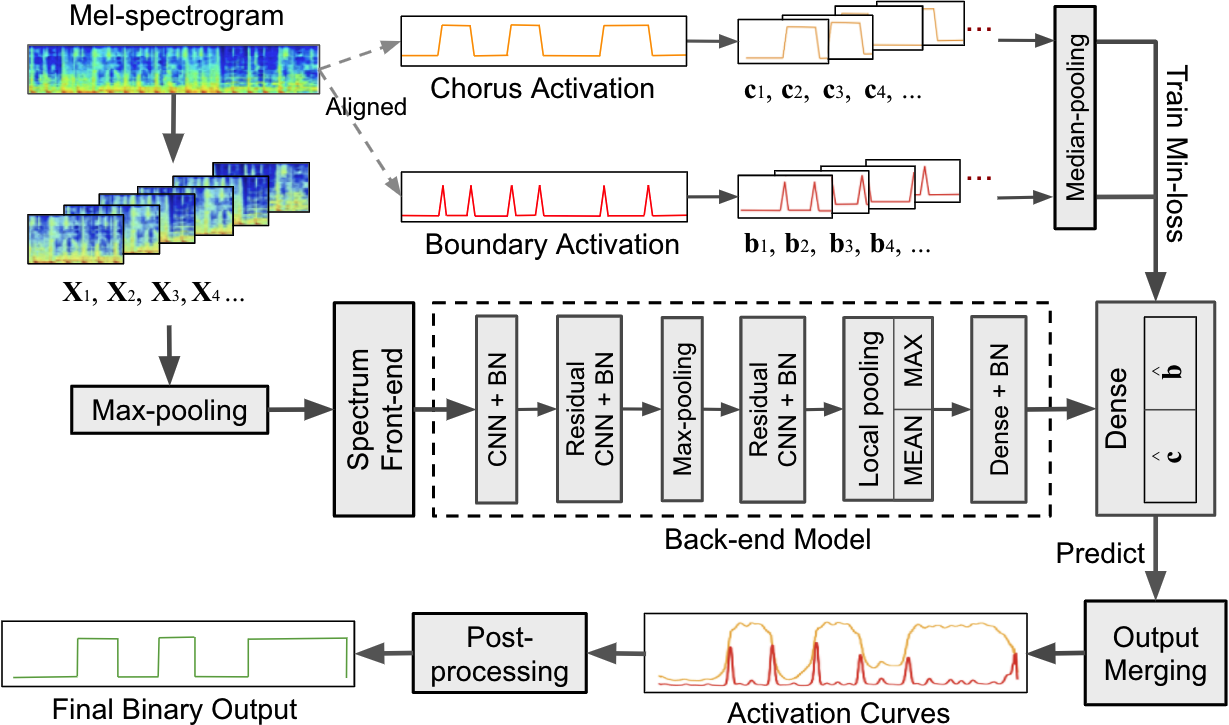}
\caption{The system diagram.}
\label{fig:SYS}
\end{figure}

This section details the three main stages of the system. The overall pipeline is illustrated in Figure~\ref{fig:SYS}.


\subsection{Feature and Label Pre-processing}
\label{sec:method_stage1}

We use the mel-spectrogram of a song as input. The model takes a window of $N$ frames (defined as a \emph{chunk}) with a hop size of $S$ frames at a time. Note that $N$ is appropriately large to allow the model to see longer contexts of the audio. 

The annotations include the starting and ending timestamps of each chorus. For each song, we create two types of target labels: a \textit{chorus activation curve} $\mathbf{c}$ and a \textit{boundary activation curve} $\mathbf{b}$.
For a song of length $L$, we define $\mathbf{c}=\left[c_1, \ldots, c_L\right]$, with $c_t=1$ if $t$ lies within a chorus section, and $c_t=0$ otherwise.
To smooth the transitions, half of a 2-second wide Hann window is used to ramp from 0 to 1 prior to the chorus onset; a similar ramp down is added after the chorus offset.
To create the boundary activation curve,
we convert each boundary instant into a ``boundary section'' of duration 0.5 seconds, and then apply the same ramp up and down.
Thus, each boundary produces a 2.5-second wide bump in $\mathbf{b}$.
We use a wider target than in~\cite{ullrich2014_ismir} to tolerate greater deviations from the true boundaries in our case, since our goal is to predict the full extent of the chorus.

In previous works~\cite{ullrich2014_ismir, grill2015_eusipco}, the system models the probability of a single target (i.e., a boundary) at the center of a chunk.
By contrast, we design the system to model the probabilities of the entire activation curve in the chunk, with each probability aligned with a frame in the mel-spectrogram. This enables the network to explicitly learn the contextual dependency from the target activation curve.
To sum up, a chunk-level training sample for the CNN is represented as $\{\mathbf{X}\in\mathbb{R}^{N\times D}, \mathbf{c}\in\mathbb{R}^{N}, \mathbf{b}\in\mathbb{R}^{N}\}$, where $\mathbf{X}$ is the mel-spectrogram using $D$ components.

\subsection{CNN-based Model}
\label{sec:method_stage2}

The model is shown in the center part of Figure~\ref{fig:SYS}. To facilitate reproducibility, we adopt the model architecture proposed in~\cite{pons2018end}, which has shown excellent performance in music classification/tagging tasks. We make three modifications to meet the requirements of our task:
First, we add a temporal max-pooling layer prior to the spectrum front-end model to sub-sample the input mel-spectrogram. We use a pool size of [6, 1] with a stride of [6, 1]. To ensure synchronization with the mel-spectrogram, we also apply median-pooling for $\mathbf{c}$ and $\mathbf{b}$ with a pool size of 6 with a stride of 6. Second, we replace the global pooling (for mean- and max-pooling over time) with a \emph{local pooling} at the penultimate layer of the back-end model. A pool size of [24, 1] and a stride of [12, 1] are used. This design serves the need to model the entire temporal activation curve. Third, we
add a final dense layer to output the chorus and boundary predictions, denoted by $\mathbf{\hat c}\in\mathbb{R}^{N/6}$ and $\mathbf{\hat b}\in\mathbb{R}^{N/6}$, respectively. All the model parameters remain the same as~\cite{pons2018end} except those mentioned above.



To achieve multi-task learning, we calculate the losses for $\mathbf{\hat c}$ and $\mathbf{\hat b}$ separately. Then, the final loss is the weighted combination: 
$\alpha \cdot \text{loss}(\mathbf{\hat c}) + (1-\alpha) \cdot \text{loss}(\mathbf{\hat b})$,
where $\alpha \in [0, 1]$ and $\text{loss}(\cdot)$ is a reduce-mean operation that averages the element-wise losses. 

\subsection{Output Merging and Post-processing}
\label{sec:method_stage3}
We obtain the chunks from a song using a large overlap (e.g. 95\%), so that during training, the model can see the labels for multiple times with multiple shifts of mel-spectrogram, which is expected to help fast convergence.
At the prediction stage, we can merge the predictions of multiple overlapping windows to improve robustness.
We take the average of the overlapped probabilities to obtain the merged activation $y[t]$ at each global time step $t \in [1, \dots, L]$ of a song, which can be formulated as follow:
\begin{equation}
    y[t] = \frac{1}{|Q(t)|}\sum_{i \in Q(t)} \hat y_i[m(i,t)],
\end{equation}
where $\left\{\hat y_i[t']\right\}$, $t'=[1,\dots,N]$ is the predicted activation of the $i$-th chunk, $m(i,t)$ is the function that maps a global time step $t$ to a local time step $t'$ for the the $i$-th chunk, and the function $Q(t)$ returns the set of chunks that are available at $t$.
For example, using 95\% overlap, $|Q(t)|$ would be 20 for most of the song, but it would ramp down to be 1 at the start and end of the song, with $|Q(1)|$ = $|Q(L)|$ = 1.
This method is used to obtain the final predicted curves for both chorus and boundary activations.


To obtain a binary prediction, we must apply some peak-picking or thresholding heuristics to the predicted activation curves.
However, we observed in our pilot study that the overall probability values can be very low for some songs that the model is less confident about. Setting a global threshold to binarize the curves could thus lead to no choruses or boundaries being detected in these songs.

To avoid this, we develop a more flexible method which makes use of the relative likelihoods of the segmented curve.
The post-processing includes three phases: (1) select top $P$ peaks from the boundary curve to partition the song into segments; (2) calculate the chorus likelihood by averaging the chorus probabilities within each segment; (3) select the top $R$ segments (by likelihood) as the choruses, and assign the others as non-choruses. 
For the first phase, we follow the peak-picking method in \cite{ullrich2014_ismir} to select boundary candidates: 
any boundary having the maximum probability within a 10-second non-overlapped window throughout the curve is kept. Each candidate is assigned a boundary score by subtracting the average of the activation curve in the past 10 and future 5 seconds.

We tailor $P$ and $R$ to the dataset, since the annotation guidelines and hence the typical number of segments for each dataset are different. For example, in Harmonix it is possible for two chorus sections to occur back-to-back with a boundary in between, but this arrangement was not possible in the In-House dataset.
Accordingly, we calculate $\theta$, the average number of choruses per 3-minutes, from the training set as prior knowledge.
We use it to set $P$ and $R$ as follows:
$P = 2.5 \times R$ and $R = 2\times d \times (\theta/180)$, where $d$ is the test song's duration in seconds. Intuitively, $d \times (\theta/180)$ is the expected number of chorus sections for a test song.
Our choice of $R$ thus reflects a strategy to slightly over-segment the song at first, which is reasonable since adjacent sections with the same predicted label will be merged.

\section{Experiments}\label{sec:experiment}

\subsection{Implementation Details}

LibROSA \cite{mcfee2015librosa} is used to extract the log-scaled mel-spectrogram with $D$ = 96 components. The waveform is resampled at 32KHz, and an FFT window of 2048 samples with 1024-sample hop size is applied. 
For segmenting chunks, we adopt a window size of $N$ = 600 frames (19.2 seconds) with a hop size of $S$ = 30. 
In our preliminary experiments, we found the value of $S$ does not significantly affect the validation accuracy when it is appropriately small (e.g. $< 50$). Since it is related to the amount of data to be processed, increasing $S$ can reduce the time complexity. 

We use $\alpha$ = 0.1, as we observed in the validation that the boundary curve is more difficult to learn. We note that our model is not sensitive to $\alpha$ when $\alpha<0.5$. Smaller $\alpha$, which emphasizes learning the boundary curve, can result in better overall results. This observation makes intuitive sense: there are far fewer positive training examples for boundary frames than for chorus frames (ratio is smaller than 0.1), so emphasizing this loss can force the model to be more careful with frames near boundaries,
which can eventually help the post-processing to make better decisions.

Our model is implemented with TensorFlow 1.15 and trained using the Adam SGD optimizer that minimizes the cross entropy loss. We use a mini-batch of 256 examples and apply batch normalization with momentum 0.9 at every layer of the network. The initial learning rate is 0.0005 and annealed by half at every 15,000 training steps.

\subsection{Experimental Settings}

We use three datasets to evaluate the proposed approach: the subset of SALAMI in the ``popular'' genre (denoted by \emph{SALAMI-pop}) \cite{smith2011design}; the Harmonix Set \cite{NietoISMIR2019} with $\theta$ = $\sim$3.7 (training sets); and an internal music collection (In-House) with $\theta$ = $\sim$2.2 (training sets).
SALAMI-pop was used for testing only, so its $\theta$ was never computed or used; the other datasets were used to conduct 4-fold cross-validation and cross-dataset evaluations.


SALAMI-pop contains 210 songs.
Since some songs are annotated twice, we treat each annotation of a song as a separate test case,  yielding 320 test cases.
For both SALAMI and Harmonix Set,
we categorized ``pre-chorus'' as non-chorus (to disentangle the build from the true chorus) and ``post-chorus'' as chorus
(since they seem more related to the chorus than to the rest of the song), and merged the segments accordingly.
The In-House dataset was compiled for the purpose of training a chorus detector. It contains 2480 full tracks covering many genres of popular music, including Chinese-pop, J-pop, K-pop, hip-hop, rock, folk, electronic, and instrumental. At least one chorus section is annotated in each track.



We study the performance of the raw chorus activation curve using the area under the ROC (AUC), and the final binary output using the pairwise F1 score,
which is the standard metric for evaluating music structure analysis~\cite{muller2015a}
and related tasks like beat/downbeat tracking~\cite{dixon2007evaluation}.


Our main proposed model is named as \textbf{Temporal} model (Section~\ref{sec:method}), because it predicts the entire temporal activation of a chunk.
We also introduce a variant, termed as \textbf{Scalar} model, that predicts a scalar chorus and boundary probability (two values) at the center of an input chunk (like in~\cite{ullrich2014_ismir,maezawa2019music}). Specifically, we set $S$ = 6, use global pooling in the back-end model, and skip the output merging stage.
To study the potential accuracy loss due to the post-processing design, we create \textbf{OracleBound}, which uses the ground-truth boundaries and uses the number of choruses for $R$ to parse the predicted chorus curve of the best-performing Temporal model. 

We compare these models to four open-source baseline systems that use existing approaches:
pychorus~\cite{Jayaram2018blog}, which is based on~\cite{goto2006chorus}, and three algorithms implemented in MSAF~\cite{nieto2016systematic}.
We optimized pychorus 
using the following heuristics: we modified it to output up to 4 top candidates (default is one);
and, when no chorus is found with an initial reference duration (15 seconds), we iteratively reduce the duration by 3 seconds until it finds a chorus. 

MSAF provides implementations of many algorithms for segmenting songs and grouping segments. None give explicit function labels like ``verse'' or ``chorus,'' but we can take the predicted segment groups as chorus candidates, and try two heuristics to guess which group represents the choruses: 
(1) \textit{Max-freq}: choose the most frequent label as the chorus, and
(2) \textit{Max-dur}: choose the segment group that covers the greatest duration of a song as the choruses.
We use the CNMF \cite{nieto2013convex}, SCluster \cite{mcfee2014analyzing}, and VMO \cite{wang2016structural} algorithms, all with default settings.
As \textit{Max-dur} consistently outperformed \textit{Max-freq} for each algorithm, we report these results only.

\subsection{Results and Discussion}\label{sec:discussion}
\begin{table}
 \begin{center}
 \begin{tabular}{|l|ccc|ccc|}
 \hline
Metric & \multicolumn{3}{c|}{AUC}  &  \multicolumn{3}{c|}{F1}  \\
\hline \hline
Model~\textbackslash~Test & HS & IH & SP & HS & IH & SP \\
\hline
Temporal-HS & \textbf{.827} & .767 & .723 & \textbf{.692} & .624 & .602 \\
Scalar-HS  & \textbf{.826} & .728 & .706 & \textbf{.688} & .597 & .585 \\
Temporal-IH & .775 & \textbf{.868} & .736 & .630 & \textbf{.668} & .596 \\
Scalar-IH  & .764 & \textbf{.860} & .735 & .616 & \textbf{.665} & .592 \\
\hline
OracleBound & - & - & - & \textbf{.738} & \textbf{.825} & .709 \\
\hline
pychorus & .629 & .585 & .557 & .466 & .378 & .330 \\
CNMF~\cite{nieto2013convex} & .570 & .524 & .525 & .479 & .367 & .416 \\
SCluster~\cite{mcfee2014analyzing} & .603 & .523 & .506 & .534 & .297 & .418 \\
VMO~\cite{wang2016structural} & .455 & .463 & .481 & .272 & .229 & .277 \\
\hline
 \end{tabular}
\end{center}
\caption{Mean score comparison on the three datasets: Harmonix Set (HS), In-House (IH), and SALAMI-pop (SP). Temporal-`X' and Scalar-`X' indicate the results of each model when trained on dataset `X'.
Results in bold were obtained using 4-fold cross-validation.
All results of the proposed models (upper 4 rows)
are significantly greater than results of the existing systems (lower 4) with p-value $< {10}^{-20}$.
}
\label{tab:summary_result}
\end{table}


The results are summarized in Table \ref{tab:summary_result}, where each value is the mean score averaged over a complete dataset. To perform cross-dataset (CD) evaluation (e.g., Temporal-HS on IH or SP), we select the best-performing model in terms of F1 from the four models trained in the cross-validation (CV) (i.e., among folds of Temporal-HS on HS), and use it to test all the songs of the other dataset (i.e., IH or SP). 

We observe, first of all, that our proposed models outperform the existing ones by a large margin: the worst of the proposed models was, on average, 0.14 greater than the best of the baseline models, for both AUC and F1.
This outcome validates our expectation that ``chorusness'' could be learned in a supervised fashion. 
Second, the Temporal models consistently outperform their Scalar counterparts; in particular, the difference between Temporal-HS and Scalar-HS is statistically significant (p-value $< {10}^{-5}$). This indicates that modeling longer contexts of the activation is a better approach, perhaps because it exploits the temporal dependency of the activation curves.
Third, although training on a dataset tends to improve performance on that dataset, we observe strong CD performance: the CD F1 scores all lie within 0.61 $\pm$ 0.03 across the three datasets, demonstrating the generalizability of our approach. Since $\theta$ is fixed by the training set, high CD performance indicates robustness to different values of $\theta$.
On the other hand, the margin between our results and the OracleBound suggests that an orthogonal approach---e.g., one based on repetition---could improve the post-processing.

\section{Conclusion and Future Work}\label{sec:conclusion}

We have presented a supervised approach to detecting choruses in music audio.
In experiments, our systems performed better than several existing ones, even when trained on other datasets.
With this promising result, we believe that more types of segment labels, such as verse, bridge and solo, can be detected with supervised learning, and with less dependence on context.
The current model is relatively simple: it only considers the local context of audio signals. It could be improved if we use features and techniques to inform it of a greater context,
such as structure features \cite{serra2012unsupervised}, recurrent architecture and attention modelling \cite{huang2018pop}.

\bibliographystyle{IEEEbib}
\bibliography{citations}

\end{document}